\documentclass[aip,reprint]{revtex4-1}
\usepackage{graphicx}% Include figure files
\usepackage{dcolumn}% Align table columns on decimal point
\usepackage{bm}% bold math
\usepackage{amsmath}

\draft % marks overfull lines with a black rule on the right

\begin{document}
\title{Space Charge Free Ultrafast Photoelectron Spectroscopy on Solids by a Narrowband Tunable Extreme Ultraviolet Light Source}
\author{Riccardo Cucini}
\email[]{Corresponding author: cucini@iom.cnr.it}
\affiliation{C.N.R. - I.O.M., Strada Statale 14, km 163.5, Trieste, Italy}
\author{Tommaso Pincelli}
\author{Giancarlo Panaccione}
\affiliation{C.N.R. - I.O.M., Strada Statale 14, km 163.5, Trieste, Italy}
\author{Damir Kopic}
\affiliation{Elettra - Sincrotrone Trieste S.C.p.A., Strada Statale 14, km 163.5, Trieste, Italy}
\affiliation{Universit$\grave{a}$ degli Studi di Trieste, Via A. Valerio 2, Trieste, Italy}
\author{Fabio Frassetto}
\affiliation{IFN-CNR, Via Trasea 7, Padova, Italy}
\author{Paolo Miotti}
\affiliation{IFN-CNR, Via Trasea 7, Padova, Italy}
\affiliation{Dipartimento di Ingegneria dell'Informazione, Universit$\grave{a}$ di Padova, via Gradenigo 6/B, Padova, Italy}
\author{Gian Marco Pierantozzi}
\affiliation{C.N.R. - I.O.M., Strada Statale 14, km 163.5, Trieste, Italy}
\author{Simone Peli}
\affiliation{Elettra - Sincrotrone Trieste S.C.p.A., Strada Statale 14, km 163.5, Trieste, Italy}
\author{Andrea Fondacaro}
\author{Aleksander De Luisa}
\affiliation{C.N.R. - I.O.M., Strada Statale 14, km 163.5, Trieste, Italy}
\author{Alessandro De Vita}
\affiliation{Dipartimento di Fisica, Universit$\grave{a}$ di Milano, via Celoria 16, Milano, Italy}
\author{Pietro Carrara}
\affiliation{Dipartimento di Fisica, Universit$\grave{a}$ di Milano, via Celoria 16, Milano, Italy}
\author{Damjan Krizmancic}
\affiliation{C.N.R. - I.O.M., Strada Statale 14, km 163.5, Trieste, Italy}
\author{Daniel T. Payne}
\affiliation{Elettra - Sincrotrone Trieste S.C.p.A., Strada Statale 14, km 163.5, Trieste, Italy}
\author{Federico Salvador}
\affiliation{C.N.R. - I.O.M., Strada Statale 14, km 163.5, Trieste, Italy}
\author{Andrea Sterzi}
\affiliation{Elettra - Sincrotrone Trieste S.C.p.A., Strada Statale 14, km 163.5, Trieste, Italy}
\author{Luca Poletto}
\affiliation{IFN-CNR, Via Trasea 7, Padova, Italy}
\author{Fulvio Parmigiani}
\affiliation{Elettra - Sincrotrone Trieste S.C.p.A., Strada Statale 14, km 163.5, Trieste, Italy}
\affiliation{Universit$\grave{a}$ degli Studi di Trieste, Via A. Valerio 2, Trieste, Italy}
\affiliation{International Faculty, University of Cologne, Albertus-Magnus-Platz, 50923 Cologne, Germany}
\author{Giorgio Rossi}
\affiliation{C.N.R. - I.O.M., Strada Statale 14, km 163.5, Trieste, Italy}
\affiliation{Dipartimento di Fisica, Universit$\grave{a}$ di Milano, via Celoria 16, Milano, Italy}
\author{Federico Cilento}
\affiliation{Elettra - Sincrotrone Trieste S.C.p.A., Strada Statale 14, km 163.5, Trieste, Italy}

\date{\today}

\begin{abstract}
%Here we report on a novel High Harmonic Generation (HHG) light source designed for space charge free ultrafast photoelectron spectroscopy (PES) on solids. The ultimate overall energy resolution achieved on a polycrystalline Au sample is $\sim$\,22 meV at 40 K. These results have been obtained at a photon energy of 16.9 eV by varying, up to 200 kHz, the photon pulses repetition rate  and the photon fluence on the sample. The bandwidth of the 16.9 eV pulses is equal to $\sim$19 meV and it sets a new benchmark for tunable narrowband HHG sources. The PES energy resolution and the photon pulse bandwidth, along with a pulse duration of $\sim$105 fs, as retrieved from time-resolved angle resolved (AR) PES experiments on Bi$_2$Se$_3$, demonstrates a space charge free photoelectric process close to Fourier transform limit conditions.
Here we report on a novel High Harmonic Generation (HHG) light source designed for space charge free ultrafast photoelectron spectroscopy (PES) on solids. The ultimate overall energy resolution achieved on a polycrystalline Au sample is $\sim$22 meV at 40 K. These results have been obtained at a photon energy of 16.9 eV with a pulse bandwidth of $\sim$19 meV, by varying, up to 200 kHz, the photon pulses repetition rate and the photon fluence on the sample. These features set a new benchmark for tunable narrowband HHG sources. By comparing the PES energy resolution and the photon pulse bandwidth with a pulse duration of $\sim$105 fs, as retrieved from time-resolved (TR) angle resolved (AR) PES experiments on Bi$_2$Se$_3$, we validate a way for a space charge free photoelectric process close to Fourier transform limit conditions for ultrafast TR-PES experiments on solids.
\end{abstract}

\maketitle %\maketitle must follow title, authors, abstract and \pacs

\section{Introduction}
\begin{figure*}[ht]
\centering
\includegraphics[width=0.9\linewidth]{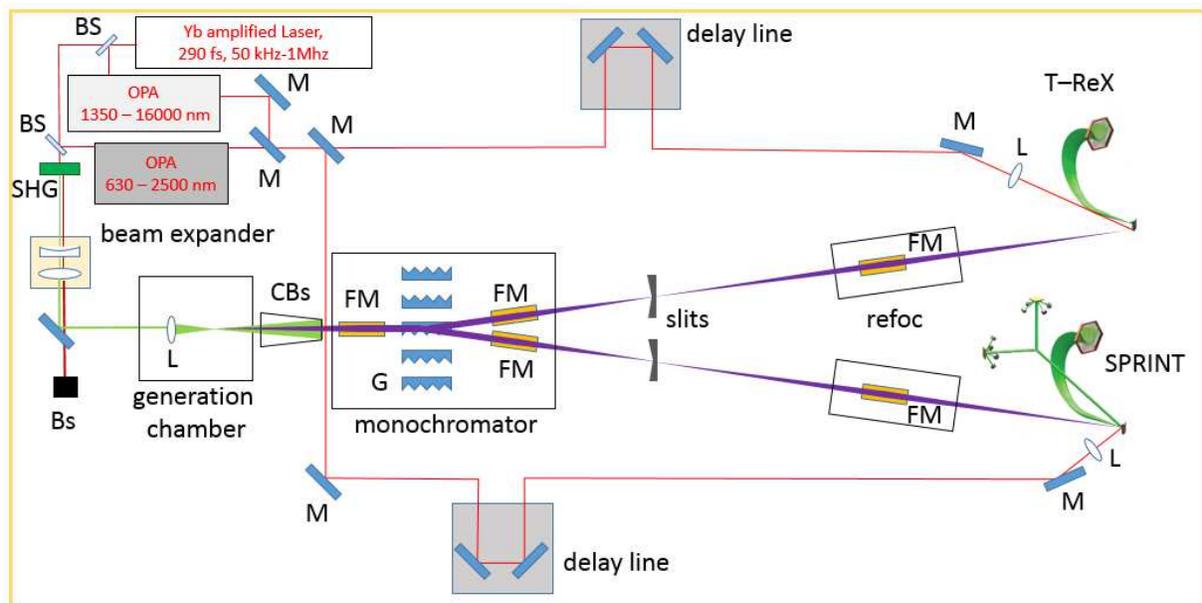}
\caption{Pump/probe setup: M, mirror; BS, beam splitter; SHG, second harmonic crystal; Bs, beam stopper; CBs, cone beam stopper; L, lens; FM focalizing mirror; G, grating.}
\label{fig1_bis}
\end{figure*}
Nowadays the expanding quest for studying the physics of complex and quantum materials under non-equilibrium conditions has prompted the development of advanced ultrafast light soources, in order to tailor specific excited states and probing their electronic transient properties. By HHG\cite{ferray,lorek,guhr,rothhardt,petersen} light sources, as well as free electron laser (FEL), it is possible to obtain radiation pulses with photon energies extending from the extreme ultraviolet (EUV) to hard X-rays, with pulse durations ranging from sub-fs to sub-ps, and  fully polarized light. However, while these sources are suitable for high peak brilliance experiments, they show severe limits for ultrafast PES on solids, where the photon density in the light pulses and the light pulses repetition rate must be controlled in order to reduce spurious effects.

PES \cite{pes1,pes2} allows to measure, under perturbative condition, the spectral function resulting from the projection of the final electronic states on the initial states of the matter. In quantum and strongly correlated materials the collective excitations and the quasi-particles interactions will affect the self-energy, hence the PES spectral function, which features unveil the effects of these many body interactions on the kinetic energy and momentum of the primary photoelectrons. By measuring the kinetic energy and the momentum of the primary electrons, the PES experiments can be extended to the reciprocal space, therefore to the measure of the band dispersion, while another degree of freedom of the photoelectron can be observed by detecting its spin.

In the last two decades ARPES and spin resolved (SP) ARPES have been extended to time domain in the sub ps regime. Although very challenging, even today, these experiments require stable photon pulses sources (in terms of energy, polarization, intensity) with duration in the range of 10-100 fs and with a controlled peak brilliance and repetition rate. Hence, the number of photons, for a fixed focal spot on the sample, has to be such that space charge effects are minimized, while compensating the limited number of photoelectron per pulse by a repetition rate as high as possible to provide the best signal-to-noise ratio.

The ideal light source for time-resolved photoelectron spectroscopy should respond to the following characteristics: i) provide pulses with 10$^3$-10$^4$ coherent photons per pulse in the energy range 6 eV-100 eV and with $\sim$ 100 fs time duration, in order to exploit favorable photoionization matrix elements, while covering the full Brillouin zone of all materials; ii) be controllable in terms of pulse power, pulse duration, and repetition rate; iii) allow tunability of the photon energy in a broad range\,\cite{dakovski,lanzara,ishida,mills,mills2,nie}; iv) have repetition rate of several tens of kHz, up to MHz\,\cite{buss,zong,corder,mills3,na,eich2,rohde,zhao,hadrich,porat,saule,li,sie,liu}, but compatible with the relaxation time of the excited states, in order to avoid the thermal effects on the sample.

In particular, space charge\cite{hellmann,oloff,oloff2} can heavily affect the primary photoelectron kinetic energy and trajectory, hence limiting the experimental resolution. This is the most challenging problem in time resolved ARPES experiments.

Here we report on a novel HHG source and setup, that lays down a new benchmark in terms of repetition rate (up to 200 kHz), photons per pulse (up to 10$^{6}$), pulse duration ($\sim$ 100 fs) and an overall energy resolution for time resolved ARPES of $\sim$\,22 meV, as measured on a polycrystalline Au at 40 K. The main characteristics of this source are summarized in Table\,\ref{tab0}. By comparing the parameters in Table\,\ref{tab0} with those of equivalent setups in the literature\,\cite{cilento,haight,petersen,puppin,rohde,buss,eich2,na,sie,frietsch,corder}, the HHG time resolved ARPES described herewith below meets the state of the art in the field.

At present, this HHG setup serves alternatively two complementary end-stations devoted to advanced PES, Spin Polarization measurements, and ARPES.
{
\begin{table}[t]
\centering
\addtolength{\tabcolsep}{3mm}
\renewcommand\arraystretch{1.5}
\begin{tabular}{cc}
\hline
\hline
Parameter & Value \\
\hline
Photon energy (eV)& 16.9\\
\hline
Rep. Rate (kHz)& 200\\
\hline
$\Delta E_{exp}$ (meV) & 22 \\
\hline
$\Delta t_{exp}$ (fs)& 300\\
\hline
$\Delta E_{probe}$ (meV)& 19\\
\hline
$\Delta t_{probe}$ (fs)& 105\\
\hline
$\Delta t_{FT}$ (fs)& 96\\
\hline
\hline
\end{tabular}
\caption{Time and energy resolution performance: $\Delta E_{exp}$, experimental energy resolution, including optical elements broadening and analyzer resolution; $\Delta t_{exp}$, experimental time resolution, measured as convolution between pump and probe pulses; $\Delta E_{probe}$, harmonic bandwidth, after deconvolution of temperature and analyzer contribution; $\Delta t_{probe}$ probe pulse time duration; $\Delta t_{FT}$, estimation of Fourier transform pulse duration, obtained considering a time-bandwidth product $\approx$0.44}
  \label{tab0}
\end{table}
}
\section{HHG SOURCE AND PHOTON BEAM OPTICS}

\subsection{The laser source}

 The source is a Yb:KGW-based integrated femtosecond laser system (PHAROS, Light Conversion), characterized by a turn-key operation and by high pulse-to-pulse stability. The system produces $\sim$300 fs pulses at 1030 nm, with a tunable repetition rate from single-shot to 1 MHz. The average power is 20 W above 50 kHz. The maximal energy-per-pulse, equal to 400 $\mu$J, is available in the 0-50 kHz interval instead. At $>$50 kHz, the energy per pulse is determined by the actual repetition rate setting; at 1 MHz, 20 $\mu$J/pulse are available. Once the fundamental repetition rate is set, the corresponding energy/pulse is provided. From this condition, and using the laser-cavity Pockels cell, a lower repetition rate can be set by pulse-picking of the pulses; this possibility preserves the energy/pulse and is particularly useful for the optimization of non-linear optical processes, since the same peak power is available but with a reduced thermal load. The repetition rate tuning flexibility has direct consequences on experiments, allowing to find the ideal compromise between (probe) signal statistics and (pump-induced) sample heating. The laser source can seed two optical parametric amplifiers (OPAs), configured for operation with 40 $\mu$J/pulse or 360 $\mu$J/pulse. They allow for a tunable output in the range 630-2600 nm (ORPHEUS HP) and in the range 1350-4500 nm (ORPHEUS ONE HP), which is also equipped with a DFG crystal to extend the output up to 16 $\mu$m.

\subsection{HHG beamline}
The HHG photon beam is propagated to two experimental end-stations designed for PES\,\cite{pincelli2016,cilento}. The beamline is made up of a generation chamber, monochromator, and refocalization chamber, and coupled with OPAs sources for time resolved pump-probe spectroscopy. The full optical path is reported in fig.\,\ref{fig1_bis}.

The generation chamber has been designed to work in the so-called tight-focusing regime, that allows to reach the 10$^{14}$ W/cm$^2$ peak power density, required to drive the HHG process\,\cite{chiang,chiang2}, also with a pulse duration of $\sim$300 fs. A detailed description is reported in the Supplementary Material.

In standard operating conditions, the HHG process is seeded by the second harmonics of the laser, at 515 nm. The frequency duplication of the laser radiation at 1030 nm is obtained with a 2 mm thick Beta Barium Borate (BBO) nonlinear crystal. The conversion efficiency into 515 nm radiation is 50$\%$ without beam focusing.

\begin{figure}[h!]
%\centering
\includegraphics[width=1.4\linewidth]{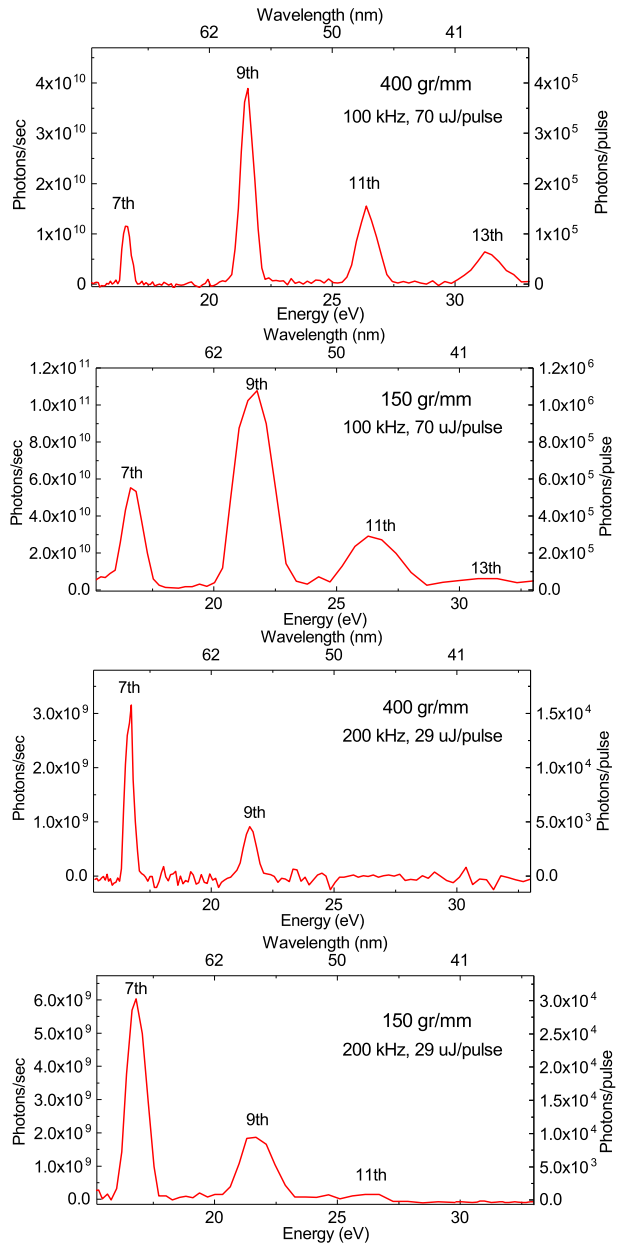}%
\caption{HHG spectra as resolved by gratings with 150 gr/mm and 400 gr/mm at repetition rate of 100 and 200 kHz. The identification of the harmonics is reported.}
\label{fig2}
\end{figure}
\subsection{Monochromator}
\begin{table*}[htbp]
\centering
\renewcommand\arraystretch{1.5}
\begin{tabular}{cccccc}
\hline
\hline
mirror & $R_{sag}$ (mm) & $R_{tan}$ (mm)& $f$ (mm) & AOI ($^\circ$) & coating\\
\hline
c1, c2, c3 & 41.9 & 8600 & 300 & 86 & gold \\
refocusing & 248 & 6813 & 1300 & 79 & gold \\
\hline
\hline
\end{tabular}
  \label{tab1}
  \caption{Toroidal mirrors specification ($R_{sag}$, sagittal radius; $R_{tan}$, tangential radius; $f$, focal length; AOI, angle of incidence)  }

\end{table*}
\begin{table*}[htbp]
\centering
\renewcommand\arraystretch{1.5}
\begin{tabular}{cccccc}
\hline
\hline
grating (gr/mm) & blazing angle ($^\circ$) & range (eV)& AOI ($^\circ$)& ruled area (mm) & Bandwidth (100 $\mu$m slit) (eV)\\
\hline
150 & 3.4 & 8-30 & 5 & 70x10 &  0.5@ 20 eV\\
200 & 4.2 & 8-30 & 5 & 70x10 & 0.5@ 20 eV\\
400 & 4.5 & 30-50 & 5 & 70x10 & 1.1@ 40 eV\\
1200 & 7 & 50-100 & 5 & 70x10 & 1.45@ 20 eV\\
\hline
\hline
\end{tabular}
  \label{tab2}
  \caption{Gratings specification}

\end{table*}
The spectral selection of a single harmonics is performed by an off-plane-mount (OPM) grating monochromator. Differently from the classical diffraction, where the grating grooves (gr) are perpendicular to the incident plane, in the off-plane geometry the incident plane is almost parallel to the grooves.  The main advantage of such a configuration is the capability to mitigate the EUV pulse-front tilt after diffraction, therefore providing a lower temporal broadening of the monochromatized pulses  \cite{poletto}. Furthermore, the off-plane mount provides much higher EUV efficiency compared to the classical one\cite{frassetto}.

The instrument design consists of two toroidal gold coated mirrors and five OPM plane gratings (see Supplementary Material). The beam is collimated by a toroidal mirror onto a specific grating, and subsequently refocused onto one of the two exit slits by a toroidal mirror with identical focal length, in order to select a single harmonic beam and to steer it in either the SPRINT (Spin Polarized Research Instrument in the Nanoscale and Time) or T-ReX (Time Resolved X-Ray spectroscopy) end-stations\,\footnote{Open access is performed through the NFFA-SPRINT and T-ReX facilities via the web site http://trieste.nffa.eu/ and http://www.elettra.eu/userarea/apbt.html, respectively.}. The mirrors are operated at equal grazing angle and unity magnification, in order to minimize the aberrations at the output. The selection of a specific grating determines which end-station is served, without further adjustments. The specifications of mirrors and gratings are reported in Table \ref{tab1} and \ref{tab2}, respectively. The gratings parameters have been designed to introduce a dispersion sufficient to isolate a single harmonics with a slit-width of $\sim$100 um (adjacent harmonics are separated by 2.4 eV with 1030 nm seed and 4.8 eV with 515 nm seed). Since the intrinsic bandwidth of the harmonics is definitely lower than the bandwidth transmitted through the monochromator, the overall energy resolution is limited by the HHG source itself. Hence, the monochromator is acting as a tunable filter that is used to select a single harmonic and filter out all the adjacent harmonics.

The use of grating gives intrinsically a pulse-front tilt due to diffraction, which introduces a temporal broadening. The measure of the EUV divergence allows to estimate this broadening. The beam divergence has been measured through the knife-edge technique, resulting in 10 mrad for harmonics in the range 16.9 eV (i.e., 7$^{th}$ harmonic of 515 nm) to 31.2 eV (i.e., 13$^{th}$ harmonic of 515 nm).
The resulting temporal broadening given by the monochromator because of the pulse-front tilt is in the range 50-200 fs, typically lower than the expected duration of the harmonics.
\subsection{Refocusing}
The refocusing chamber (see Supplementary Material) is equipped with a toroidal mirror (see Table \ref{tab1} for specifications), which images the monochromatized EUV beam spot at the slits position onto the sample plane, with 1:1 ratio. A silver square mirror is placed in the refocusing chamber, to direct the pump beam on the sample. The pump beam forms a $\approx$1 degree angle with respect to the EUV probe pulse. The mirror is mounted on a piezo-mount to set precisely the pump-probe beam overlap. A second square mirror, mounted off-center on a stepper motor, can be inserted in the beam path to send the quasi-collinear pump-probe beams out of the vacuum chamber. This possibility is used to optimize the beams focus (note that the pump beam is focused by a lens with focal length f = 1.5 m, placed before the entrance window on the refocusing chamber) and to roughly determine the time-zero condition. To this aim, the pump beam and the zero-order probe are sent on a fast photodiode recorded by a 4 GS/s oscilloscope. In this way, time-zero is pre-determined with a few ps uncertainty and finally found directly in the photoemission experiments.
\begin{figure*}[t]
\centering
\includegraphics[width=0.7\linewidth]{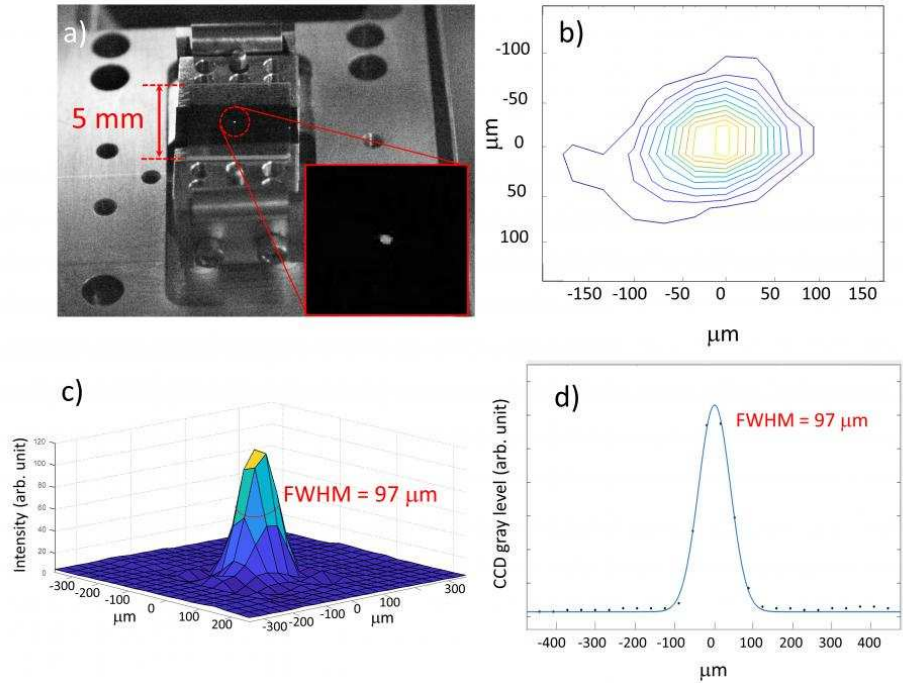}%
\caption{EUV focal spot. a) Focal spot on the YAG screen; b) contour plot of the focal spot; c) 3D profile; d) horizontal profile. The blue line represents the Gaussian fit.}
\label{fig3}
\end{figure*}
\section{Results}

Fig.\,\ref{fig2} reports the harmonics spectrum as measured by low energy gratings (150 and 400 gr/mm). Spectra were recorded under the same generation conditions, namely, the HHG process is driven by the second-harmonic-generation (SHG) of the laser output (70 $\mu$J/pulse at 100 kHz and 29 $\mu$J/pulse at 200 kHz; the additional energy losses after the SHG crystal are due to the optical elements along the beam path), in Argon (4 bar at nozzle entrance).
The spectra have been measured by scanning the monochromator with 0.05 degrees steps. The measurement of the photon flux at the output of the monochromator, after passing the slit with an aperture of 100$\,\mu$m, has been performed with an X-ray photodiode, connected to an acquisition board. The conversion between current and number of photons has been obtained by using an analogue photodiode calibrated at a synchrotron radiation facility. Each point has been acquired for 5 seconds. The measurements were performed at full power, at two different repetition rates. The number of photons per second measured in this condition is $\approx$0.4-1.1x10$^{11}$ at 100 kHz, and $\approx$1-6x10$^{9}$ at 200 kHz.

Harmonics generation in Neon with 1030 nm has been also tested, using a dedicated grating with 1200 gr/mm, generating up to the $63^{rd}$ harmonics (not reported), with a two orders of magnitude lower photon flux.

The EUV focal spot is measured with a Yttrium Aluminium Garnet (YAG) crystal placed at the sample position. The EUV-beam fluorescence is recorded by a Charge-Coupled Device (CCD) camera. The horizontal profile of the EUV spot is fitted by a Gaussian function. The FWHM is 97 $\mu$m, as shown in Fig.\,\ref{fig3}).

The HHG beamline feeds two end-stations optimised for complementary PES experiments:  a) The T-ReX end-station, hosting a SPECS electron analyzer equipped with a Delay Line Detector (DLD), optimized for time-resolved ARPES experiments. The T-ReX end-station features a six-degrees of freedom motorized cryomanipulator, which hosts the sample during experiments; details of the experimental setup can be found in\,\cite{cilento}(http://www.elettra.trieste.it/labs/t-rex.html); b) the SPRINT end-station, where standard characterization techniques for surface science experiments (LEED, AES, ion bombardment) and a Scienta SES 2002 electron energy analyser (EA) are available\,\cite{pincelli2016,pincelli2017}. The experimental chamber also hosts  a Vectorial Twin Mott detector set-up\,\cite{petrov} for the measurement of spin polarization of the full, or partialized, quantum yield from the sample\,\cite{pincelli2016}, and a Helium discharge lamp for reference spectra at 21.22 eV with 1 meV  bandwidth and analyzer energy resolution calibration. Selected results obtained in commissioning experiments are reported in the following subsections.

\subsection{Space Charge Mitigation}
\begin{figure}[t]
\centering
\includegraphics[width=1\linewidth]{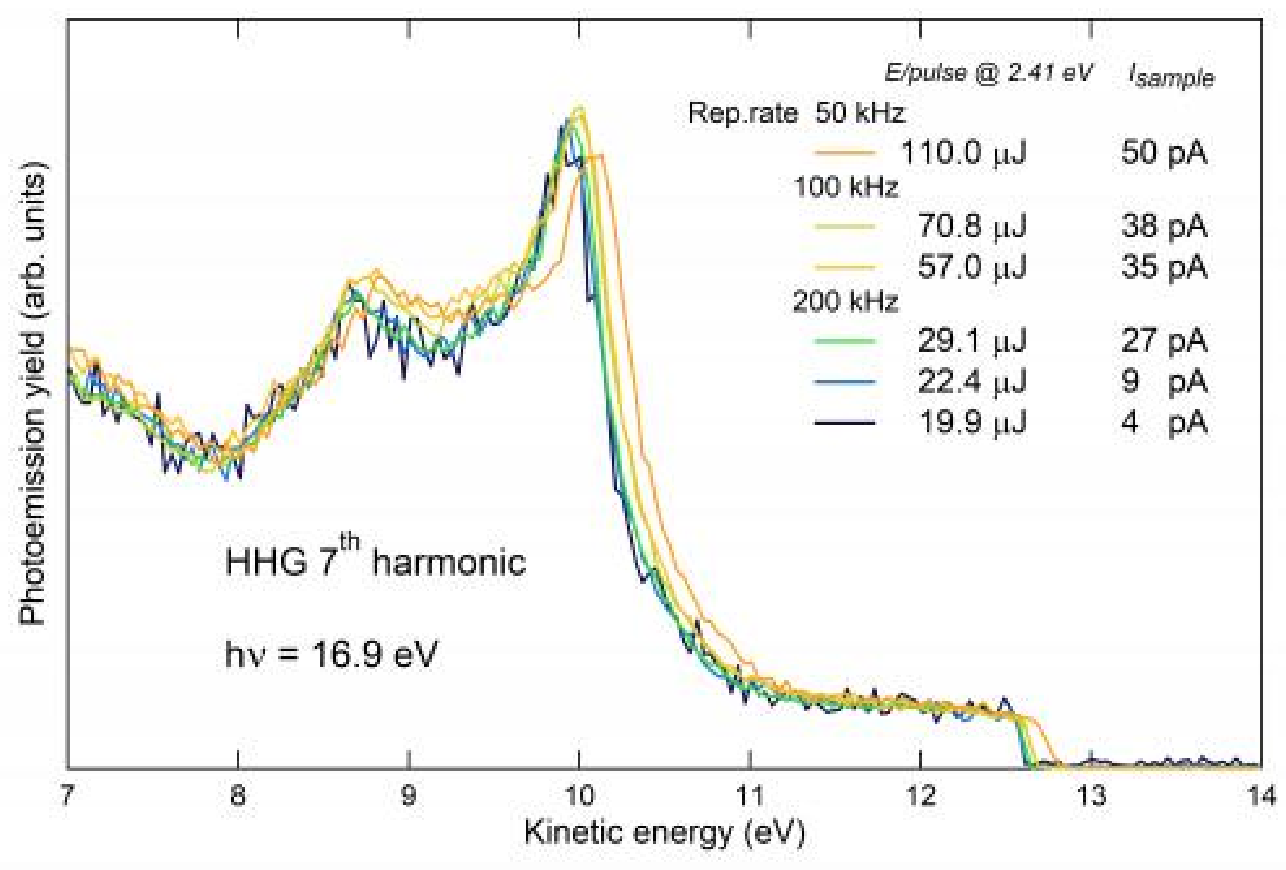}\\
\includegraphics[width=1\linewidth]{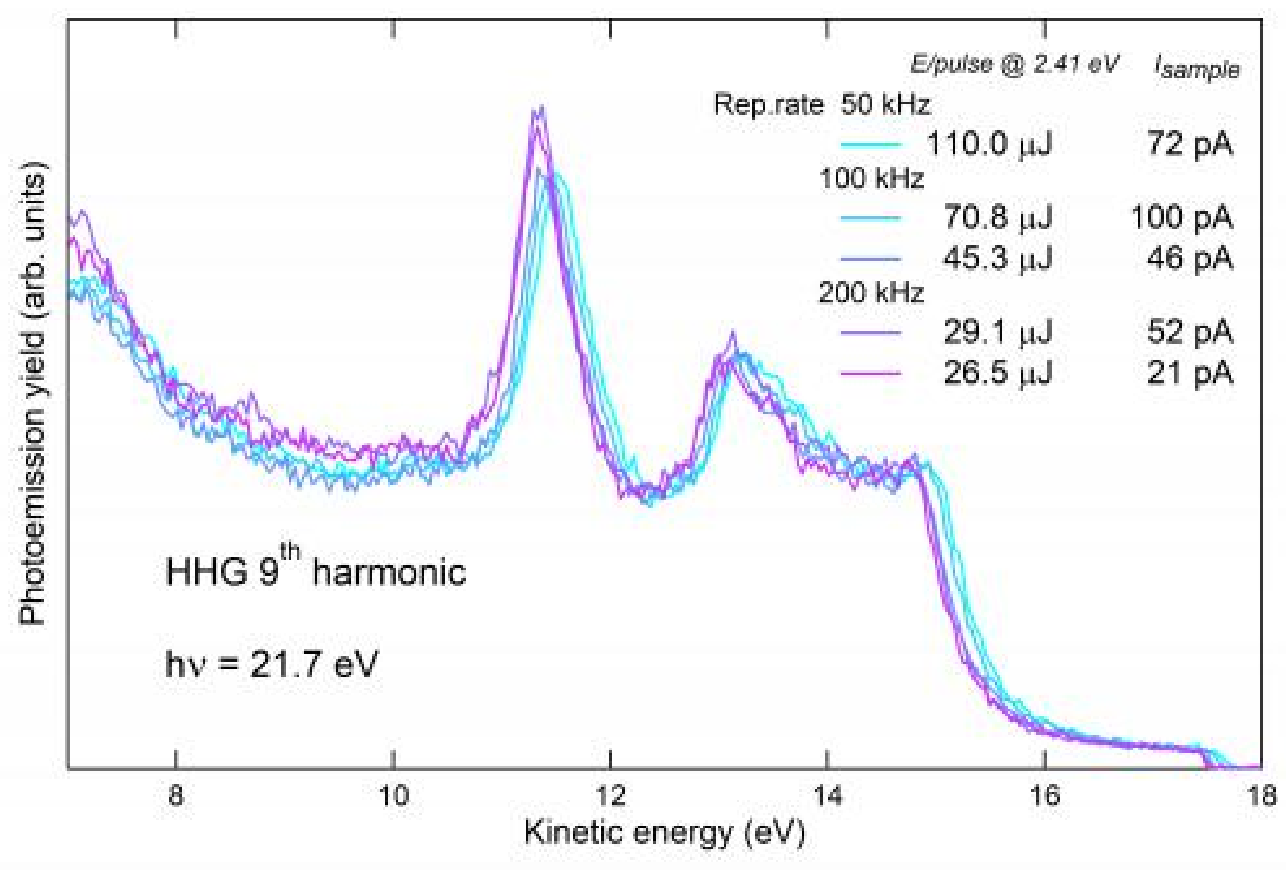}%
\caption{Angle integrated photoemission spectra of polycrystalline gold foil at T = 40 K, with 16.9 eV photon energy (7$^{th}$ HHG harmonic, upper panel) and  21.7 eV photon energy (9$^{th}$ HHG harmonic, lower panel), changing the energy per pulse of the driving laser from 110 $\mu$J to 19.9 $\mu$J; for each spectrum the total electron yield (I$_{sample}$) is reported as a measure of the total power impinging on the sample. Spectra are rescaled so as to have the same height of the Fermi step.}
\label{fig4}
\end{figure}
When ultrashort Ultraviolet (UV) pulses exceeding 10$^5$ photons per pulse impinge on a sample surface, a large number of photoelectrons are created. The emitted electron cloud in front of the sample has the shape of a thin disk, hence the electrons strongly interact with each other. This results in the fact that fast electrons are pushed ahead and slow electrons are retarded, producing an energy-shifted and momentum-distorted photoemission spectrum when more than one electron per laser shot is emitted\cite{zhou,fognini}.
The typical strategies to mitigate space charge effects in photoemission are the reduction of the photon flux, or the reduction of photon density on the sample surface by increasing the spot size. However, these workarounds are far from optimal, as they lead to a reduction of the statistics of the measurement. This is a critical problem when dealing with low repetition rate pulsed UV or X-ray sources, like the present FEL (up to a hundred hertz) and the few kHz-range HHG sources.

The high repetition rate operation (up to 200 kHz) of our setup overcomes these problems. Indeed, high repetition rate operation makes it possible to maintain a high flux with a moderate intensity of the individual pulses, mitigating space charge.

The laser source used provides constant power (20 W) above 50 kHz operation. Hence one can reduce the energy per pulse available for high harmonic generation by increasing the repetition rate. However, due to the high nonlinearity of the generation process, the harmonic beam power is not constant. In addition, also the harmonics energy cutoff is decreased. This sets an upper limit to the maximal repetition rate for each harmonics.

In order to demonstrate the possibility to mitigate space charge effects, we measured the valence band of a clean surface of a polycrystalline gold foil (Au displays a flat 6s-like Density of State (DOS) across the Fermi level, suitable for the estimation of the energy broadening and shift induced by space charge) while tuning the generation conditions. In particular, we choose three different repetition rates (50 kHz, 100 kHz, 200 kHz), corresponding to three different energies per pulse at 515 nm (200 $\mu$J, 100 $\mu$J, 50 $\mu$J, respectively) after the BBO. Due to losses from optical components, the energy per pulse into the generation chambers is 110 $\mu$J, 70 $\mu$J, 29 $\mu$J, respectively. The energy per pulse at fixed repetition rate can be also changed by an external attenuator (we report measurements down to 19.9 $\mu$J at 200 kHz). It allows to analyze the evolution of the photoemission spectra as a function of the seed energy-per-pulse at a given repetition rate.

\begin{figure*}[t]
\centering
\includegraphics[width=1\linewidth]{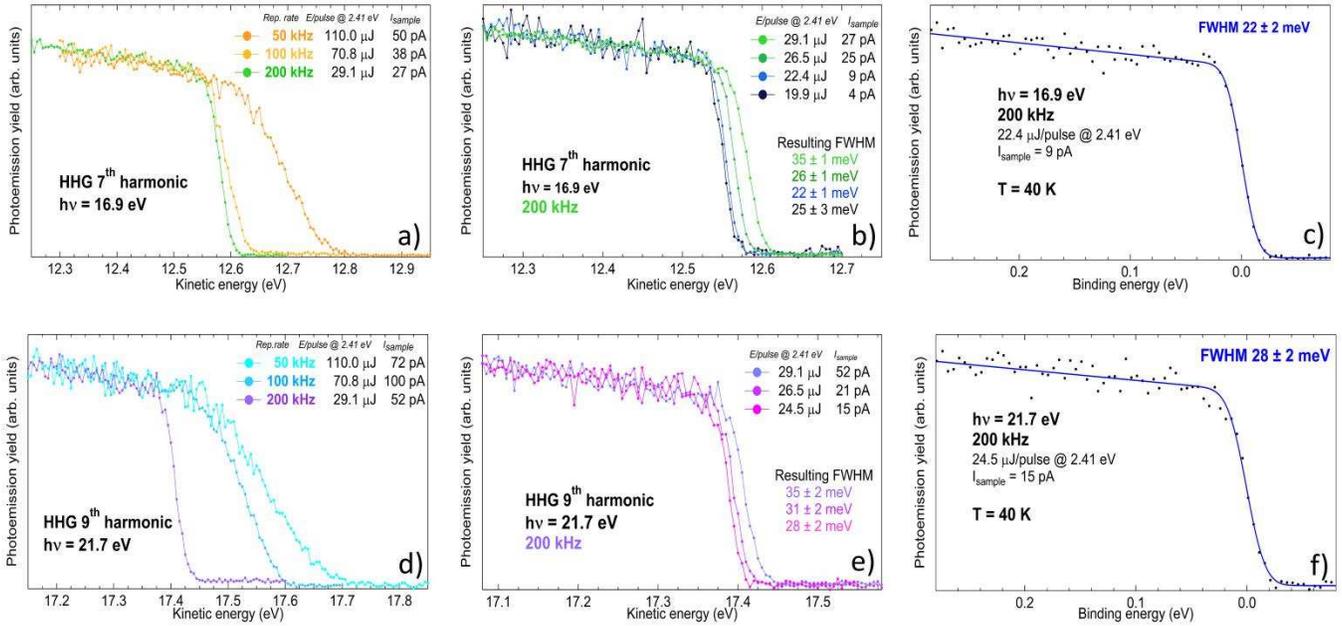}%
\caption{Fermi edge measured at 40 K for different laser repetition rates and photon energies. a) Fermi edge at 16.9 eV (7$^{th}$ harmonic) measured at 50, 100, 200 kHz repetition rate; b) Fermi edge at 16.9 eV, 200 kHz, varying the energy per pulse of the driving laser; c) 7$^{th}$ harmonic energy bandwidth measurement.  Square black dots: experimental data. Solid blue line: fitting curve used to extract the FWHM; d) Fermi edge at 21.7 eV (9$^{th}$ harmonic) measured at 50, 100, 200 kHz repetition rate; e) Fermi edge at 21.7 eV, 200 kHz, varying the energy per pulse of the driving laser; f) 9$^{th}$ harmonic energy bandwidth measurement.}
\label{fig5}
\end{figure*}

The angle integrated photoemission spectra obtained with the 7$^{th}$ and the 9$^{th}$ harmonic are displayed in Fig.\,\ref{fig4} as a function of the energy per pulse of the driving laser. The spectra are measured at a temperature T = 40 K. For each spectrum, the total electron yield is measured by recording the drain current with a picoammeter (Keithley 6482), which connects the sample to the electrical ground. As the amount of space charge is reduced, the spectra shifts towards lower kinetic energy and the Fermi edge (E$_F$) becomes steeper. This feature is showed in detail in Fig.\,\ref{fig5}. Panel a) shows a zoom of the Fermi edge on gold measured with the 7$^{th}$ harmonic, changing the repetition rate from 50 kHz (orange curve) to 200 kHz (green curve): a strong reduction of the broadening and the shift of the Fermi step is achieved.

Panel b shows the spectra measured at 200 kHz upon further reducing the energy-per-pulse by means of attenuation of the driving laser: some very small space charge effects are actually still present at maximum fluence (green curve), and the space charge is totally removed when the total electron yield is below 20 pA (blue and black curves), corresponding to around 600 photoelectrons per pulse. Panels d) and e) show spectra measured with the 9$^{th}$ harmonic, where a similar trend can be observed.

\subsection{Harmonics Energy Bandwidth}\label{bandwidth}
Once the space-charge related effects have been controlled, a reliable measurement of the best attainable energy resolution can be performed by investigating the Fermi edge of polycrystalline Au at low temperature.

The Fermi edges measured with the 7$^{th}$ harmonic at 16.9 eV and the 9$^{th}$ harmonic at 21.7 eV at the SPRINT end-station (analyzer pass energy 5 eV, entrance slit 0.5 mm) are displayed in Fig.\,\ref{fig5} panel c) and f), respectively. The fitting curve (blue line) is a convolution of the Fermi function $f(E_K)=(e^{\frac{E_K-E_F}{k_B T}}+1)^{-1}$, with a Gaussian, accounting for the instrumental resolution, including the source photon bandwidth. While the temperature is fixed and known by independent measurement, the free parameters of the fit are $E_F$ and the FWHM of the Gaussian itself,  and the slope and intercept of the line mimicking the DOS.
The resulting FWHM is 22$\pm$2 meV at 16.9 eV and 28$\pm$2 meV at 21.7 eV. This value can be decomposed as the sum in quadrature of the source and the detector contributions\,\cite{torelli}, as well as other instrumental broadenings (electronic noise):
$FWHM^2=\bigtriangleup E^2_{source}+\bigtriangleup E^2_{detector}+ \bigtriangleup E^2_{other}$
According to the formula $\bigtriangleup E_{detector}=(W/2R_0) E_p$ (W = 0.5 mm, R$_0$ = 200 mm, E$_p$ = 5 eV), the expected value of the detector resolution in the configuration of this measurement is 6.25 meV (to this value one should add the term $\alpha^2/4 E_p$ which accounts for the angular spread $\alpha$ of the electrons transmitted into the hemispheres)\,\cite{ibach}. We have evaluated directly the detector resolution by measuring the Fermi edge at liquid nitrogen temperature with a known source (He I lamp, 21.22 eV) as a function of the Pass Energy, yielding  $\bigtriangleup E_{detector}=(0.0019 \pm 0.0006) \times E_P$; since in our case $Ep= 5$ eV, the result is $\bigtriangleup E_{detector}=9.5 \pm 3.0$ meV, in good agreement with the expected value, considering the presence of the angular term.
Such value is negligible with respect to the total FHWM retrieved by the fit in Fig.\,\ref{fig5}, panel c) and f). Hence, we conclude that the reported values are the upper limit for the overall energy resolution measured with the 7$^{th}$ and 9$^{th}$ harmonic. Most importantly, in Fig.\,\ref{fig5}, panel b) and e), we report the overall energy resolution obtained with presence of space charge, for both the harmonics at 200 kHz. Also in the worst condition, the overall energy resolution never exceeds 35 meV.

\subsection{Harmonics Duration}
\begin{figure}[t]
\centering
\includegraphics[width=0.8\linewidth]{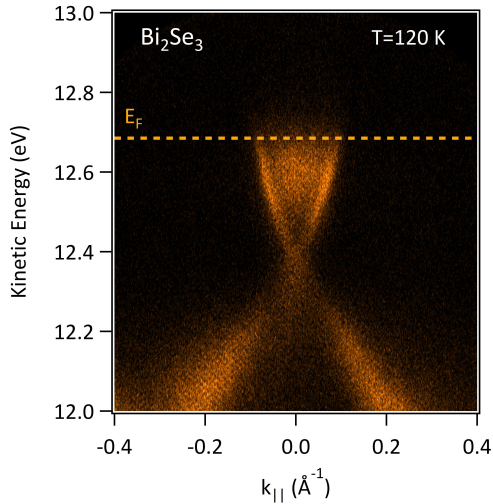}%
\caption{ARPES map acquired on Bismuth Selenide at T=120 K, and h$\nu$=16.9 eV (harmonics 7$^{th}$).}
\label{fig7}
\end{figure}
\begin{figure}[t]
\centering
\includegraphics[width=0.8\linewidth]{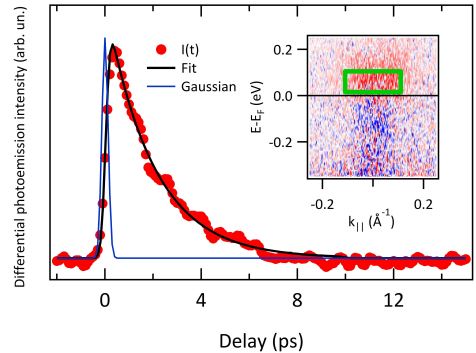}%
\caption{Electron dynamics acquired in the pump-probe experiment. Photoemission intensity has been averaged in the green box overlaid on the ARPES map reported as inset, which shows the differential ARPES intensity collected at the delay t=300 fs. The black line in the main panel is the fit to the data (see main text for details). Blue line is the Gaussian representing the pump-probe cross-correlation, as retrieved by the fitting routine.}
\label{fig8}
\end{figure}
In this section we report on the results obtained by Angle-Resolved PES (ARPES) at equilibrium and out-of-equilibrium, using the EUV high-harmonics as a probe. The experiments have been performed at the T-ReX end-station equipped with a SPECS Phoibos 225 hemispherical electron analyzer, using the topological insulator Bismuth Selenide (Bi$_2$Se$_3$, provided by HQ Graphene) as a reference sample. The sample temperature was set to T=120 K. Fig.\,\ref{fig7} shows the ARPES map acquired on the Bi$_2$Se$_3$ sample in s-polarization, using the 7$^{th}$ harmonics (h$\nu$=16.9 eV) as a probe. The pass-energy of the analyzer was set to 15 eV (the entrance-slit width is 0.5 mm), and the repetition rate of the laser source was 100 kHz. The total acquisition time was $\approx$10 minutes.
The value of 100 kHz was set for reducing the average pump power. The intensity of the EUV harmonics can be set to avoid space charge effect by reducing the intensity of the laser pulse before the generation. Under these conditions, space-charge is minimized, and a good count rate is achieved, as it is appreciated from the quality of the ARPES map. The out-of-equilibrium experiment was performed under the same experimental conditions, the only difference being the analyzer slit width, set to 2 mm to reduce the acquisition time. As a pump, we used the fundamental of the laser source, at 1030 nm. The beam was focused on the sample through a f=1.5 m focal length lens, down to a spot size of 300$\pm$10 $\mu$m. The pump-probe spatial overlap was set by superimposing the two beams on a Ce:YAG scintillator, that converts EUV radiation in visible photons (at $\approx$560 nm) that are imaged by a complementary metal-oxide semiconductor (CMOS) camera. In the experiment, the pump fluence was set to 150$\pm$30 $\mu$J/cm$^2$. The results of the pump-probe experiment are reported in Fig.\,\ref{fig8}. In the pump-probe scan, the step of the delay-line was set to 66.67 fs, which is below the expected time resolution. The integration time for data reported in Fig.\,\ref{fig8} is 2 hours. The inset of Fig.\,\ref{fig8} shows the differential ARPES map computed by subtracting a map acquired before excitation (t=-500 fs) from the map acquired at t=+300 fs. In this way, the effect of the pump excitation is evident. The trace reported in Fig.\,\ref{fig8} shows the electron dynamics extracted by averaging the ARPES intensity in the green box overlaid to the ARPES map reported as inset. The electron dynamics has been analyzed with a single exponential decay convoluted to a Gaussian (reported in Fig.\,\ref{fig8} as a solid blue line) representing the pump-probe cross correlation. The only free parameters of the fit (solid black line) are the intensity I and time-constant $\tau$ of the exponential decay, and the Gaussian FWHM, $\sigma$. We obtain $\tau =2 \pm 0.1$ ps and $\sigma =300 \pm 30$ fs.

The duration of the pump-pulse has been measured independently with an auto-correlator (APE Berlin PulseCheck) in the last portion of the pump beam path, and resulted $\sigma_{pump} = 280 \pm 5$ fs. With this value, we determine an upper limit for the XUV probe FWHM, after deconvolution of $\sigma_{pump}$ from $\sigma$. The result is: $\sigma_{probe}=105 \pm 45$ fs, which is reasonable if we consider the natural pulse shortening obtained both in the SHG process (the FWHM of the 515 nm seed beam is $230 \pm 5$ fs, as measured with the auto-correlator) and in the subsequent HHG process. The value we provide constitutes an upper limit for the harmonics duration, since a non-negligible rise-time in the TR-ARPES signal measured on Bi$_2$Se$_3$ is likely present. For Gaussian pulses the minimum bandwidth required to sustain a pulse-duration of 105 fs is $\approx$17 meV. By comparing this value with the upper limit of the energy bandwidth as determined by photoemission measurements, we can conclude that we are close to the transform-limit condition.
\section{Concluding Remarks}

%In this work we have addressed one of the major issues in TR-ARPES and TR-PES by mitigating the space charge effects convoyed by the photoelectric process in solids.
%
%These effects, chiefly severe in time resolved photoelectric experiments, significantly deplete the overall energy and momentum resolution.
%
%Here we demonstrate that by controlling the photon beam parameters, the photon fluence on the sample and the pulse repetition rate it is possible to perform space-charge free photoemission with ultrashort EUV photon pulses.
%
%For the momentum integrated Fermi edge of polycrystalline Au at 40 K we achieved an ultimate overall energy resolution of $\sim$ 22 meV, with a pulse duration of $\sim$ 105 fs.
%
%This resolution, when compared to the HHG pulse bandwidth, i.e. $\sim$19 meV, and duration, $\sim$ 105 fs, demonstrates that time resolved PES experiments can be performed under Fourier transform limit conditions.
%
%As a direct consequence, the HHG source, photon beam optics and photon transport system, as reported in this work, open the way to TR-PES and TR-ARPES experiments where the optimal trade-off between time and energy resolution is achieved, while the full Brillouin zone of crystalline solids can be accessed.
In this work we have addressed one of the major issues in TR-ARPES and TR-PES by mitigating the space charge effects convoyed by the photoelectric process in solids. These effects, chiefly severe in time resolved photoelectric experiments, significantly deplete the overall energy and momentum resolution.

Here we demonstrate that by controlling the photon beam parameters, the photon fluence on the sample and the pulse repetition rate it is possible to perform space-charge free photoemission with ultrashort EUV photon pulses.

For the momentum integrated Fermi edge of polycrystalline Au at 40 K we achieved an ultimate overall energy resolution of $\sim$22 meV. This resolution, close to the HHG pulse bandwidth, i.e.$\sim$19 meV, when compared with a photon pulse duration of $\sim$105 fs, demonstrates that time resolved PES experiments can be performed under Fourier transform limit conditions.

As a direct consequence, the HHG source, the photon beam optics and the photon transport system, as reported in this work, open the way to TR-PES and TR-ARPES experiments where the optimal trade-off between time and energy resolution is achieved, while the full Brillouin zone of crystalline solids can be accessed.

\section*{supplementary material}
See Supplementary Material for a detailed description of the HHG beamline.

\begin{acknowledgments}
The authors would like to thank C. Spezzani for his suggestions on HHG generation, F. Sirotti for advise on the spin polarization set-up, and G. Cautero and all the Elettra Sincrotrone electronic workshop for the development of acquisition system. This work has been partially performed in the framework of the Nanoscience Foundry and Fine Analysis (NFFA-MIUR Italy Progetti Internazionali) facility.
\end{acknowledgments}
\providecommand{\noopsort}[1]{}\providecommand{\singleletter}[1]{#1}%

\end{document}

\title{Supplementary material:\\Space Charge Free Ultrafast Photoelectron Spectroscopy on Solids by a Narrowband Tunable Extreme Ultraviolet Light Source}
\author{Riccardo Cucini}
\email[]{Corresponding author: cucini@iom.cnr.it}
\affiliation{C.N.R. - I.O.M., Strada Statale 14, km 163.5, Trieste, Italy}
\author{Tommaso Pincelli}
\author{Giancarlo Panaccione}
\affiliation{C.N.R. - I.O.M., Strada Statale 14, km 163.5, Trieste, Italy}
\author{Damir Kopic}
\affiliation{Elettra - Sincrotrone Trieste S.C.p.A., Strada Statale 14, km 163.5, Trieste, Italy}
\affiliation{Universit$\grave{a}$ degli Studi di Trieste, Via A. Valerio 2, Trieste, Italy}
\author{Fabio Frassetto}
\affiliation{IFN-CNR, Via Trasea 7, Padova, Italy}
\author{Paolo Miotti}
\affiliation{IFN-CNR, Via Trasea 7, Padova, Italy}
\affiliation{Dipartimento di Ingegneria dell'Informazione, Universit$\grave{a}$ di Padova, via Gradenigo 6/B, Padova, Italy}
\author{Gian Marco Pierantozzi}
\affiliation{C.N.R. - I.O.M., Strada Statale 14, km 163.5, Trieste, Italy}
\author{Simone Peli}
\affiliation{Elettra - Sincrotrone Trieste S.C.p.A., Strada Statale 14, km 163.5, Trieste, Italy}
\author{Andrea Fondacaro}
\author{Aleksander De Luisa}
\affiliation{C.N.R. - I.O.M., Strada Statale 14, km 163.5, Trieste, Italy}
\author{Alessandro De Vita}
\affiliation{Dipartimento di Fisica, Universit$\grave{a}$ di Milano, via Celoria 16, Milano, Italy}
\author{Pietro Carrara}
\affiliation{Dipartimento di Fisica, Universit$\grave{a}$ di Milano, via Celoria 16, Milano, Italy}
\author{Damjan Krizmancic}
\affiliation{C.N.R. - I.O.M., Strada Statale 14, km 163.5, Trieste, Italy}
\author{Daniel T. Payne}
\affiliation{Elettra - Sincrotrone Trieste S.C.p.A., Strada Statale 14, km 163.5, Trieste, Italy}
\author{Federico Salvador}
\affiliation{C.N.R. - I.O.M., Strada Statale 14, km 163.5, Trieste, Italy}
\author{Andrea Sterzi}
\affiliation{Elettra - Sincrotrone Trieste S.C.p.A., Strada Statale 14, km 163.5, Trieste, Italy}
\author{Luca Poletto}
\affiliation{IFN-CNR, Via Trasea 7, Padova, Italy}
\author{Fulvio Parmigiani}
\affiliation{Elettra - Sincrotrone Trieste S.C.p.A., Strada Statale 14, km 163.5, Trieste, Italy}
\affiliation{Universit$\grave{a}$ degli Studi di Trieste, Via A. Valerio 2, Trieste, Italy}
\affiliation{International Faculty, University of Cologne, Albertus-Magnus-Platz, 50923 Cologne, Germany}
\author{Giorgio Rossi}
\affiliation{C.N.R. - I.O.M., Strada Statale 14, km 163.5, Trieste, Italy}
\affiliation{Dipartimento di Fisica, Universit$\grave{a}$ di Milano, via Celoria 16, Milano, Italy}
\author{Federico Cilento}
\affiliation{Elettra - Sincrotrone Trieste S.C.p.A., Strada Statale 14, km 163.5, Trieste, Italy}

\date{\today}

\maketitle %\maketitle must follow title, authors, abstract and \pacs

\section{Methods}

\begin{figure*}[t]
\centering
\includegraphics[width=0.6\linewidth]{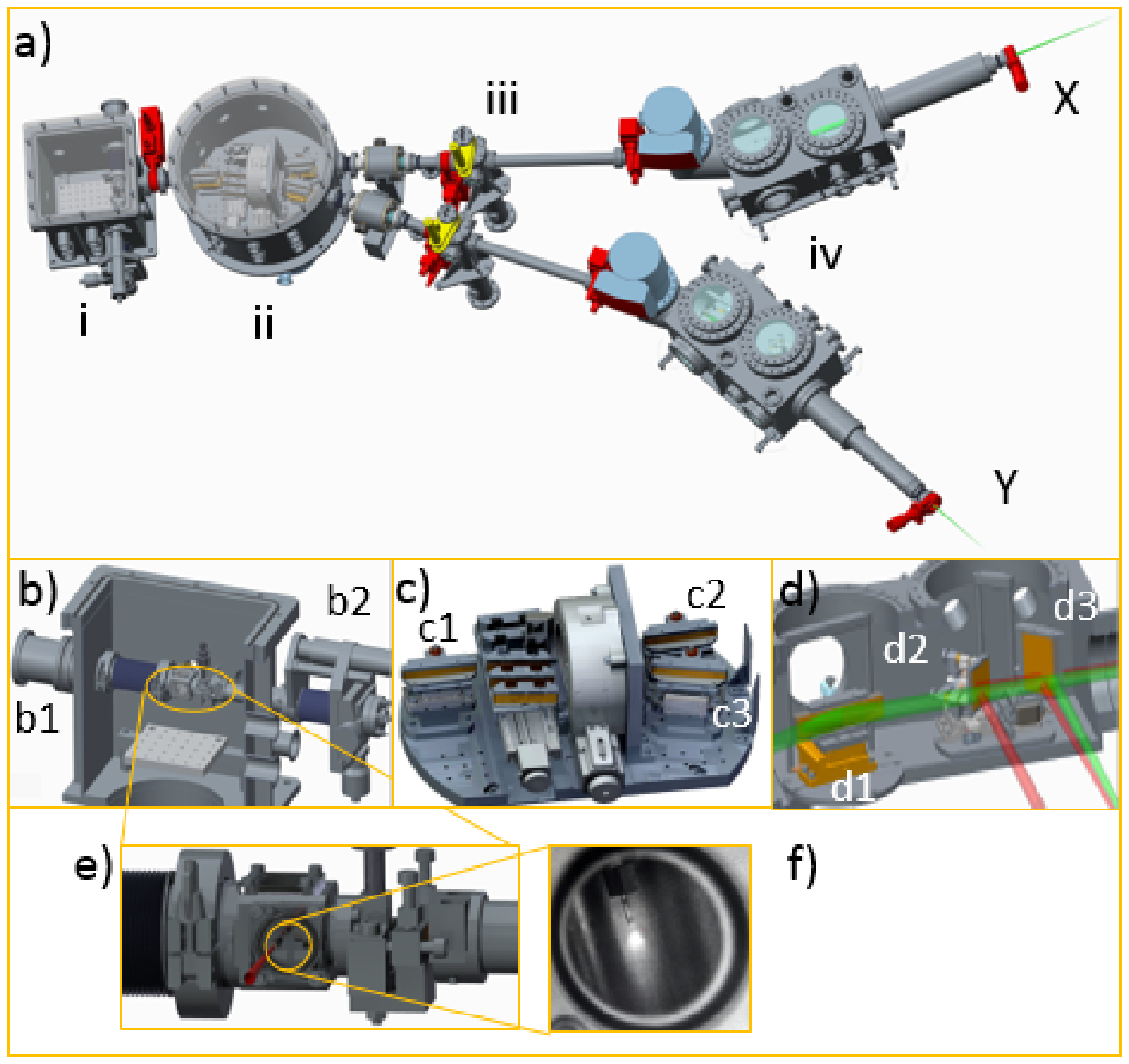}
\caption{a) overview of the two-branch beamline. The three-dimensional drawing of the HHG-beamline is shown. The overall extension of the vacuum system is $\approx$2.5 m. From left to right: i) generation chamber; ii) monochromator with exit slits; iii) photodiode chambers; iv) refocusing chambers; X: T-ReX branch; Y: SPRINT branch; b) detail of the generation chamber: b1, pumping port; b2 nozzle manipulator; c) the optics in the monochromator chamber are shown. The beam is collimated by a first toroidal mirror (c1) onto a specific grating, and subsequently refocused onto one of  the two exit slits by a second  toroidal mirror (c2 or c3, depending on which branch is in use) with identical focal length. The five diffraction gratings are mounted in the so called  ``off-plane'' geometry in order to mitigate temporal broadening of EUV pulses; d) detail of the refocusing chamber: d1, refocusing mirror; d2, pump mirror, which allows to send an external laser beam, overlapped with the HHG beam at the sample position, for pump/probe measurements; d3, metallic mirror, which allows to send outside the chamber the pump beam and the fundamental radiation from laser, for time overlapping; e) generation region, with gas nozzle and the laser entrance and exit; f) plasma plume from laser focusing on the gas jet.}
\label{fig1}
\end{figure*}

\subsection{HHG beamline}

The main application of the system described is the generation of high-order EUV harmonics in inert gases (Ar and Ne in particular). The HHG photon beam is propagated by a differentially pumped UHV beamline shown in fig.\,\ref{fig1}a) to two experimental end-stations designed for PES\,\cite{pincelli2016,cilento}. These end-stations have complementary features, while sharing the possibility of in-situ preparation and treatment of surfaces and the use of docking stations able to receive, in UHV, samples grown elsewhere. The first section of the beamline is devoted to harmonics generation, and comprises a monochromator to select a specific harmonics and  send the monochromatic beam to one of the two end-stations. The EUV beam is finally focused to a small spot size ($<$100 $\mu$m Full Width Half Maximun (FWHM)) at the sample position of the end-stations. The full optical path for pump/probe spectroscopies is reported in fig.\,\ref{fig1}g).
\subsection{Harmonic generation}
The HHG process is a highly-nonlinear optical effect emerging when the intensity of the laser light electric field is comparable to the atomic bond strength of a medium, most often a gas\,\cite{ferray}. This effect, known since more than 30 years, leads to the simultaneous generation of a number of odd harmonics of the seed photon energy, with almost constant intensity over a wide (plateau) energy region. It is usually described within the so-called three-step-model: tunnel ionization, free acceleration and recombination (recollision)\,\cite{krause}.

The generation chamber (i in Fig.\,\ref{fig1}a)) is shown in detail in Fig.\,\ref{fig1}b) and e). In order to maintain a good vacuum level without overloading the main turbomolecular pump, a second chamber is installed around the gas nozzle, consisting in a sharp glass tip having 70 $\mu$m internal diameter. This chamber is directly connected via an in-vacuum feedthrough to a 140 m$^3$/h primary pump. The only apertures of the inner chamber towards the main chamber are two adjustable holes for the beam entrance and exit. This solution allows to routinely apply a gas pressure of several (4-6) bars at the gas nozzle input, while maintaining a base pressure of $\approx 10^{-5}$ mbar in the main chamber. The gas nozzle is connected to a translation stage, used to optimize the nozzle position with respect to the laser beam.

In the standard operating conditions, the HHG process is seeded by the second harmonics of the laser, at 515 nm. The frequency duplication of the laser radiation at 1030 nm is obtained with a 2 mm thick Beta Barium Borate (BBO) nonlinear crystal. The conversion efficiency into 515 nm radiation is 50$\%$ without beam focusing. The beam is then focussed on the gas jet with a 10 cm focal length lens. We estimate the spot-size at the focus to be 10$\pm$2 $\mu$m.
In order to dump the laser radiation used for generation, a cone-shaped beam stopper is inserted between the generation chamber and the monochromator. A hole of 1.5 mm on the vertex of the cone guarantees the complete transmission of the harmonics, while limiting the seed beam transmission to only $\approx$1$\%$ of the input power. In this way, a longer lifetime of mirrors and gratings is expected. The beam dump also introduces a high vacuum impedance, reducing the pressure rise in the following vacuum chambers due to the generation gas.

\providecommand{\noopsort}[1]{}\providecommand{\singleletter}[1]{#1}%